%% file: Main.tex
\documentclass[acmsmall]{acmart}

\input{tool}

\setcopyright{cc}
\setcctype{by-nc-nd}
\acmDOI{10.1145/3832273}
\acmYear{2026}
\acmJournal{PACMSE}
\acmVolume{3}
\acmNumber{ISSTA}
\acmArticle{ISSTA182}
\acmMonth{10}
\acmSubmissionID{issta26main-p2097-p}
\received{2026-01-30}
\received[accepted]{2026-06-25}

\begin{document}

\title[EventSpec: Defining and Detecting Event-Semantic Issues\ldots]{EventSpec: Defining and Detecting Event-Semantic Issues in Blockchain Ecosystems}

\author{Yixuan Liu}
\orcid{0009-0006-2255-7901}
\affiliation{\institution{Nanyang Technological University}\city{Singapore}\country{Singapore}}
\email{liuy0255@e.ntu.edu.sg}

\author{Yuxin Dong}
\orcid{0009-0003-6239-4690}
\affiliation{\institution{Peking University}\city{Beijing}\country{China}}
\email{dongyuxin@stu.pku.edu.cn}

\author{Ye Liu}
\orcid{0000-0001-6709-3721}
\affiliation{\institution{Beijing Institute of Technology}\city{Beijing}\country{China}}
\email{ye.liu@bit.edu.cn}

\author{Yin Wu}
\orcid{0009-0001-6583-3703}
\affiliation{\institution{Xi'an Jiaotong University}\city{Xi'an}\country{China}}
\email{wuyin@stu.xjtu.edu.cn}

\author{Chengxuan Zhang}
\orcid{0000-0003-0186-9560}
\affiliation{\institution{Nanyang Technological University}\city{Singapore}\country{Singapore}}
\email{chengxua001@e.ntu.edu.sg}

\author{Xiapu Luo}
\orcid{0000-0002-9082-3208}
\affiliation{\institution{Hong Kong Polytechnic University}\city{Hong Kong}\country{China}}
\email{csxluo@comp.polyu.edu.hk}

\author{Yi Li}
\orcid{0000-0003-4562-8208}
\affiliation{\institution{Nanyang Technological University}\city{Singapore}\country{Singapore}}
\email{yi\_li@ntu.edu.sg}

\renewcommand{\shortauthors}{Yixuan Liu, Yuxin Dong, Ye Liu, Yin Wu, Chengxuan Zhang, Xiapu Luo, and Yi Li}

\input{Sections/0_Abstract}


\maketitle
\input{Sections/1_Introduction}
\input{Sections/2_Background}
\input{Sections/3_Defects}
\input{Sections/4_Methodology}
\input{Sections/5_Evaluation}
\input{Sections/6_Related}
\input{Sections/7_Discussion}
\input{Sections/8_Conclusion}
\input{Sections/9_DataAvailability}

\begin{acks}
We thank the following three external experts for their valuable feedback in
reviewing and validating our taxonomy: Peiyu Wang from CertiK, Weibo Wang from
ExVul Security Inc., and Dr. Jianzhong Su, a Research Fellow at Nanyang
Technological University. This research is supported by the
Singapore Ministry of Education Academic Research Fund Tier 2
(T2EP20224-0003), Academic Research Fund Tier 1 (RG12/23), and the Nanyang
Technological University Centre for Computational Technologies in Finance
(NTU-CCTF). Any opinions, findings, and conclusions or recommendations
expressed in this material are those of the author(s) and do not necessarily
reflect the views of MOE and NTU-CCTF.
\end{acks}

\bibliographystyle{ACM-Reference-Format}
\bibliography{bibliography}

\end{document}

%% file: tool.tex
\usepackage{subcaption}
\usepackage{multirow}
\usepackage[linesnumbered,ruled,vlined]{algorithm2e}
\usepackage{listings}
\usepackage{textcomp}

\usepackage{amsfonts}
\usepackage{xspace}
\usepackage{makecell}

\usepackage{siunitx}

\usepackage[capitalize]{cleveref}
\crefname{section}{Sect.}{Sects.}
\Crefname{section}{Section}{Sections}
\crefname{definition}{Def.}{Defs.}
\Crefname{definition}{Definition}{Definitions}
\crefname{algorithm}{Alg.}{Algs.}
\Crefname{algorithm}{Algorithm}{Algorithms}
\definecolor{verylightgray}{rgb}{.97,.97,.97}
\lstdefinelanguage{Solidity}{
  keywords=[1]{anonymous, assembly, assert, balance, break, call, callcode, case, catch, class, constant, continue, constructor, contract, debugger, default, delegatecall, delete, do, else, emit, event, experimental, export, external, false, finally, for, function, gas, if, implements, import, in, indexed, instanceof, interface, internal, is, length, library, log0, log1, log2, log3, log4, memory, modifier, new, payable, pragma, private, protected, public, pure, push, require, return, returns, revert, selfdestruct, send, solidity, storage, struct, suicide, super, switch, then, this, throw, transfer, true, try, typeof, using, value, view, while, with, addmod, ecrecover, keccak256, mulmod, ripemd160, sha256, sha3}, 
  keywordstyle=[1]\color{blue}\bfseries,
  keywords=[2]{address, bool, byte, bytes, bytes1, bytes2, bytes3, bytes4, bytes5, bytes6, bytes7, bytes8, bytes9, bytes10, bytes11, bytes12, bytes13, bytes14, bytes15, bytes16, bytes17, bytes18, bytes19, bytes20, bytes21, bytes22, bytes23, bytes24, bytes25, bytes26, bytes27, bytes28, bytes29, bytes30, bytes31, bytes32, enum, int, int8, int16, int24, int32, int40, int48, int56, int64, int72, int80, int88, int96, int104, int112, int120, int128, int136, int144, int152, int160, int168, int176, int184, int192, int200, int208, int216, int224, int232, int240, int248, int256, mapping, string, uint, uint8, uint16, uint24, uint32, uint40, uint48, uint56, uint64, uint72, uint80, uint88, uint96, uint104, uint112, uint120, uint128, uint136, uint144, uint152, uint160, uint168, uint176, uint184, uint192, uint200, uint208, uint216, uint224, uint232, uint240, uint248, uint256, var, void, ether, finney, szabo, wei, days, hours, minutes, seconds, weeks, years},  
  keywordstyle=[2]\color{teal}\bfseries,
  keywords=[3]{block, blockhash, coinbase, difficulty, gaslimit, number, timestamp, msg, data, gas, sender, sig, value, now, tx, gasprice, origin},  
  keywordstyle=[3]\color{violet}\bfseries,
  identifierstyle=\color{black},
  sensitive=false,
  comment=[l]{//},
  morecomment=[s]{/*}{*/},
  commentstyle=\color{gray}\ttfamily,
  stringstyle=\color{red}\ttfamily,
  morestring=[b]',
  morestring=[b]"
}
\newcommand{\tool}[0]{\textsc{EventSpec}\xspace}
\newcommand{\TargetPoolSize}{6{,}617 }

\newif\ifrevisionmode
\revisionmodefalse
\newcommand{\revised}[1]{\ifrevisionmode{\color{blue}#1}\else#1\fi}

%% file: Sections/0_Abstract.tex
\begin{abstract}
In recent years, smart contracts have become the backbone of decentralized applications (DApps), and
off-chain systems such as bridges, wallets, and indexers rely heavily on event logs to track
contract execution and state changes. However, the Ethereum Virtual Machine (EVM) does not
validate or enforce event semantics, so logs can diverge from on-chain state, misleading
off-chain systems into accepting incorrect state transitions. Existing smart contract vulnerability detection tools focus on logic bugs, with limited support for detecting event-semantic defects. To address this gap, we collect audit reports and incident cases and apply
open card sorting to define five classes of event-semantic defects: event collision, state-event
mismatch, unauthorized event emission, event emission mismatch, and event parameter mismatch. We propose \tool,
which infers event specifications from a contract corpus via behavior inference and
semantic-constraint extraction and applies differential
checking to identify event-semantic defects in target contracts. We run \tool on \TargetPoolSize real-world contracts and evaluate detection effectiveness based on manually labeled results; \tool achieves an overall comprehensive precision of 90.17\%. We further provide an off-chain evaluation harness that reproduces two off-chain attack vectors on any EVM-compatible chain: event origin confusion caused by unintended emitters and event–state desynchronization where events lack matching state updates. Using this harness, we demonstrate the feasibility of these attacks on bridge relayers, blockchain explorers, and NFT marketplaces, and report six wallet issues, four of which were confirmed (including a \$600 bounty), with two remaining pending.
\end{abstract}

%% file: Sections/1_Introduction.tex
\section{Introduction}
\label{Sec:Introduction}

Blockchain is a decentralized digital ledger technology that securely records and verifies transactions across multiple computers, ensuring data integrity and transparency~\cite{bitcoin}. Building on blockchain technology, Decentralized Applications (DApps) have emerged as a key innovation~\cite{buterin2014next}. These applications run on smart contracts deployed across various blockchain networks, such as Ethereum, Polygon, and Binance Smart Chain (BSC). These contracts encode on-chain logic for managing state and digital assets, while a wide range of off-chain systems, such as bridges, wallets, explorers, and indexing services, interact with them to present user-facing state and functionality. 

Smart contracts can emit \textbf{events}, which the EVM records on-chain as \textbf{transaction logs}. These logs provide a structured interface for off-chain consumption, and off-chain systems rely on them to track on-chain execution and maintain user-facing state. As of 2026-01-17, 97.9\% of contract-involved successful Ethereum transactions emit at least one log~\cite{dune}, indicating that logs constitute a primary execution signal on Ethereum.

However, the EVM does not enforce the semantic correctness of logs. Smart contracts can emit events that omit, misstate, or imitate underlying state transitions. When off-chain systems treat logs as authoritative without validating emitters or checking contract state, misleading logs can trigger asset theft, fraud, and interoperability failures across applications.

Event-semantic issues arise when emitted events do not accurately summarize the underlying
execution or when their origin and parameters are insufficiently validated. These issues
include missing or extra emissions, parameter inconsistencies, events emitted under
insufficient guards and so on. They are amplified by off-chain listeners assuming that logs from expected or known contracts imply valid state transitions, and that event parameters are safe to consume without validating the underlying state. Prior work only examined event misuse in specific domains, including cross-chain bridges~\cite{Xscope,liao2024smartaxe,wang2024xguard}, NFT ownership~\cite{sleepminting}, code-documentation inconsistencies~\cite{doccon}, and token behaviors~\cite{TokenScope}. However, a cross-application view of event semantics and their off-chain attack vectors remains underexplored. A unified taxonomy with detection methodology are still missing.

To close this gap, we conduct an empirical study of real-world incident reports and audit findings to derive a taxonomy of event-semantic defects and their associated off-chain attack vectors. Through open card sorting, we identify five types of event-semantic defects and abstract two recurring off-chain attack vectors that characterize how misleading logs propagate into off-chain systems. Based on this taxonomy, we develop \tool, a bytecode-oriented detector that constructs a corpus-derived event-emission specification and performs differential checking on target contracts. The specification captures both event layout properties (e.g., signatures and indexed positions) and semantic constraints (e.g., parameter origins and guard or binding signals), enabling \tool to flag deviations with explainable evidence, augmented by symbolic equality checks for identifying event-collision anomalies. In addition, \tool includes an off-chain evaluation harness for examining flaws in off-chain log consumption and validation.

We run \tool on \TargetPoolSize real-world contracts and evaluate detection effectiveness based on manually labeled results. \tool achieves an overall comprehensive precision of 90.17\%. We further use the off-chain harness to reproduce the two attack vectors in a controlled environment and identify issues in bridge relayers, explorers, wallets, and NFT marketplaces. Finally, we discuss mitigation strategies for both smart contracts and off-chain systems to reduce such risks.

In short, we make the following contributions in this paper:
\begin{itemize}
\item \textbf{Event-Semantic Taxonomy.}
We systematically analyze incidents and audit reports and define five types of event-semantic issues together with two off-chain attack vectors.
\item \textbf{Corpus-Driven Detection.}
We propose \tool, a bytecode-oriented framework that learns event and function semantic profiles from a corpus and performs differential checking on target contracts with explainable findings.
\item \textbf{Empirical Evaluation.}
We run \tool on \TargetPoolSize real-world contracts and evaluate detection effectiveness based on manually labeled results; \tool achieves an overall comprehensive precision of 90.17\%.
\item \textbf{Real-World Feasibility.}
We implement an off-chain harness to assess the two attack vectors in real systems and report confirmed issues across bridge relayers, explorers, wallets, and NFT marketplaces (including a \$600 bounty), alongside on-chain UEM findings.
\end{itemize}

%% file: Sections/2_Background.tex
\section{Background}
\label{sec:Background}
\subsection{Smart Contract Events and Transaction Logs in EVM-Based Blockchains}
Smart contracts, written in high-level languages such as Solidity~\cite{Solidity} and Vyper~\cite{Vyper}, are compiled into bytecode and executed on the EVM~\cite{EVM}. EVM-based blockchains support a wide range of functionalities, from simple value transfers to complex decentralized applications (DApps)~\cite{DApp}. During execution, contracts may emit \textbf{events}, which the EVM records as \textbf{transaction logs} in the transaction receipt. Logs provide a structured format for off-chain consumption. Each log contains the address of the emitting contract, a set of topics, and a data field. For non-anonymous events, \texttt{topic0} stores the \texttt{keccak256} hash of the event signature, while \emph{indexed} parameters are stored in subsequent topic slots and non-indexed parameters are stored in the data field. Logs are emitted via the \texttt{LOG0--LOG4} opcodes and are not part of the contract state, making them accessible only to off-chain consumers. For anonymous events, the event signature hash is omitted from \texttt{topic0}, and indexed parameters directly occupy the topic slots.

If a transaction reverts, all logs produced during its execution are discarded. A \textbf{transaction} is a user-submitted action that specifies a \textbf{sender} address, a target address, an optional value transfer, and input data encoding the invoked function and parameters. The sender signs the transaction, while the \textbf{emitter} refers to the contract that produces events during execution. Off-chain \textbf{listeners}, including DApps, wallets, and monitoring services, subscribe to events through RPC interfaces and process matching logs to react to on-chain activities. The EVM does not enforce semantic consistency between emitted logs and the underlying contract state, so off-chain systems must validate event emitters and relevant state transitions when correctness or security is required.

\subsection{Token Standards}

In EVM-based blockchains, tokens are implemented as smart contracts following standards like Ethereum Request for Comments 20 (\textit{ERC-20}) and \textit{ERC-721}~\cite{EIP20,EIP721}, which define interfaces and events to ensure DApp compatibility. For example, ERC-20 specifies \texttt{Transfer} and \texttt{Approval} events for tracking token transfers and approvals, while ERC-721 adopts similar events for unique NFTs. These ERC-defined events enable external systems like wallets and exchanges to monitor token activities in real time, updating user balances or ownership records.

\subsection{Off-Chain Modules}
Many DApps rely on off-chain components (e.g., relayers, indexers, and monitors) that observe on-chain transactions and events and react in real time. Cross-chain bridges are a typical case: events emitted on the source chain trigger off-chain processing and subsequent operations on the destination chain. In a standard design, a bridge contract emits a deposit event when assets are locked; an off-chain relayer monitors logs, extracts parameters, and invokes the destination-chain contract to mint or release assets. Events thus act as the primary coordination signal for cross-chain transfers. If relayers accept logs without validating the emitter or corresponding state changes, misleading events can trigger incorrect off-chain actions. In practice, many listeners assume that logs from a trusted address uniquely represent the intended action and that event parameters accurately reflect state transitions, making direct state validation unnecessary. These assumptions improve performance but create the attack surface.

%% file: Sections/3_Defects.tex
\section{Understanding Event-Semantic Issues and Attack Vectors}
\label{sec:defects}
\begin{figure}[t]
\centering
\small
\setlength{\tabcolsep}{4pt}
\renewcommand{\arraystretch}{1.1}
\begin{tabular}{|
    >{\raggedright\arraybackslash}p{2.8cm} |
    >{\raggedright\arraybackslash}p{9.6cm} |
}
\hline
\textbf{Issue Name} &
FidoMeta -- Missing functionality
\\ \hline

\textbf{Severity \& Status} &
\textcolor{red}{Critical} \ \& \ Fixed (\texttt{66f43e8})
\\ \hline

\textbf{Description} &
Emits \texttt{Transfer} events without updating balances.
\newline
\textbf{Location:} \texttt{FidoMeta.sol} (\texttt{\_burn}, \texttt{\_mint}, ctor)
\\ \hline

\textbf{Impact} &
May cause double-spending or incorrect reward accounting.
\\ \hline

\textbf{Bug Code Snippet} &
{\footnotesize
\begin{lstlisting}[language=Solidity]
function _mint(address account, uint256 amount) internal onlyOwner {
    require(account != address(0));
    require(totalSupply() + amount <= cap());
    _total = _total.add(amount);
    emit Transfer(address(0), account, amount);
}
\end{lstlisting}
}
\\ \hline

\textbf{Recommendation} &
Ensure consistency between state updates and emitted events.
\\ \hline
\end{tabular}
\caption{An example card generated from an event-semantic issue}
\Description{A sample defect card for an inconsistent Transfer event. The card records the source, severity, project, issue description, impact, vulnerable code, and recommended state-update fix.}
\label{fig:card-example}
\end{figure}

We conduct an empirical study based on real-world security audit reports and attack incidents from EVM-compatible blockchains to define and classify event-semantic defects and their associated off-chain attack vectors.

\subsection{Data Collection}
\subsubsection{Security Incidents}
We collected publicly reported security incidents involving event-semantic defects by searching incident reports released by major blockchain security companies using the keywords
\emph{event} and \emph{log}. The sources include public reports from SlowMist~\cite{SlowMist}, BlockSec~\cite{BlockSec}, PeckShield~\cite{PeckShield}, and
CertiK~\cite{CertiK}. From these sources, we initially identified 46 security incidents.

\subsubsection{Audit Reports}
To complement incident data and cover a broader range of event-semantic defects, we analyzed smart contract audit reports collected by FORGE~\cite{forgeicse2026}. Starting from the full set of 6,454 audit reports, we applied keyword-based filtering to identify findings related to \emph{event} and \emph{log} usage, resulting in 1,354 security findings from 1,059 reports.

\subsection{Data Analysis}
\subsubsection{Manual Filtering}
\label{sec:manual-filter}
To ensure relevance, two researchers independently reviewed all incidents and audit findings and resolved disagreements through discussion.
\revised{Inter-rater agreement on the relevance/exclusion classification was measured using \emph{Cohen's Kappa coefficient} ($\kappa$)~\cite{cohen1960coefficient}, which quantifies inter-rater agreement corrected for chance; we obtained $\kappa = 0.89$, which under the Landis--Koch interpretation~\cite{landis1977measurement} ($\kappa \geq 0.81$) indicates \emph{almost perfect} agreement. Disagreements were resolved by joint discussion, with two additional authors as adjudicators for unresolved cases.}
We excluded items not fundamentally tied to event semantics (e.g., generic phishing unrelated to event logs, front-end manipulation, key leakage, or unrelated bugs such as arithmetic errors or reentrancy), and retained cases where event usage, parameters, or event-state semantics directly affected off-chain interpretation or decision-making.
\revised{All severity levels were retained, since event-semantic consequences depend on how off-chain consumers interpret the logs: a low-severity finding can still produce critical impact via a vulnerable bridge or wallet.}
The final dataset contains 1{,}020 audit reports with 1{,}294 findings and 11 security incidents\revised{, eight of which caused $\sim$\$115.6M in direct financial losses (e.g., Qubit Bridge \$80M~\cite{Qubit}, pNetwork \$13M~\cite{pNetwork}, Meter Bridge \$4.4M~\cite{Meter}), evidencing the substantial real-world harm caused by event-semantic defects, while the remaining three are event-driven phishing incidents (e.g., spoofed \texttt{Transfer} logs misleading wallets and explorers) with undisclosed losses}.

\subsubsection{Open Card Sorting}
We applied open card sorting~\cite{spencer2009cardsorting}, consistent with prior empirical smart contract studies~\cite{copy-paste,SSR}. Each confirmed incident or audit finding was converted into a card containing many fields, such as project name, issue description, and root cause captured in contract logic or code snippets. Two researchers, both with more than three years of smart-contract security research experience, conducted a two-round procedure: (i) jointly analyzing a random 40\% sample to derive initial categories, and (ii) independently analyzing the remaining 60\% using the derived categories, resolving disagreements through discussion.
\revised{Inter-rater agreement on the independent 60\% subset was $\kappa = 0.95$ (\emph{almost perfect}), and disagreements were resolved using the same protocol as in Section~\ref{sec:manual-filter}.}
We excluded cards lacking sufficient evidence to attribute root causes (e.g., ambiguous contexts). \revised{Among the 1{,}294 retained audit findings, Event Emission Mismatch is the largest category (about 85\%), reflecting that missing or redundant event emissions are the most prevalent event-semantic issues in smart contracts.}

\Cref{fig:card-example} illustrates an example card generated from an event-semantic issue. In this case, several functions emit \texttt{Transfer} events without updating the corresponding state of user balances. Although the emitted events indicate token transfers, the underlying contract state remains unchanged. This mismatch between events and state can mislead off-chain systems that rely on events for accounting or validation, potentially resulting in incorrect balance tracking or double-spending risks. Based on root cause analysis, this issue is classified as a \emph{State-Event Mismatch}.

\subsection{Event-Semantic Issues in Smart Contracts}
\label{sec:event-issues}
\begin{table}[!t]
  \centering
  \small
  \caption{Definitions of the Five Event-Semantic Issues}
  \Description{Event-semantic issue taxonomy. The five rows define event collision, state-event mismatch, unauthorized event emission, event emission mismatch, and event parameter mismatch.}
  \label{table:Definition}
  \setlength\tabcolsep{1.5pt}
  \resizebox{\columnwidth}{!}{
    \begin{tabular}{>{\raggedright\arraybackslash}m{0.3\linewidth} | >{\raggedright\arraybackslash}m{0.69\linewidth}}
      \hline
      \multicolumn{1}{l|}{\textbf{Issue}} & \multicolumn{1}{l}{\textbf{Definition}} \\
      \hline
      \textit{Event Collision} & Same event signature with overlapping parameter domains, obscuring origin. \\
      \hline
      \textit{State-Event Mismatch} & \revised{Emitted logs are inconsistent with the on-chain state transition they claim to summarize.} \\
      \hline
      \textit{Unauthorized Event Emission} & Privileged events emitted without access control or checks. \\
      \hline
      \textit{Event Emission Mismatch} & \revised{Event-emission behavior is wrong: missing, unexpected, extra, duplicated, misordered, or wrong-typed events.} \\
      \hline
      \textit{Event Parameter Mismatch} & \revised{The emitted event has the expected type, but its parameter schema, indexing, ordering, or value binding is wrong.} \\
      \hline
    \end{tabular}
  }
\end{table}
Our card-sorting analysis yielded five types of event-semantic defects, each defined in \Cref{table:Definition}.
\revised{To validate the taxonomy beyond inter-author agreement, we consulted three independent smart-contract security experts\label{sec:external-experts}: a senior researcher with a dozen top-venue blockchain-security publications, a CertiK auditing partner with seven years of smart-contract auditing experience, and the founder of ExVul Security~\cite{ExVul} with a decade of cybersecurity practice. After reviewing each category for meaningfulness, mutual exclusivity, and exhaustiveness, all three experts endorsed the five categories with no missing class identified, and contributed targeted refinements reflected in the per-category definitions.}

\subsubsection{Event Collision (EC)}
Event collision arises when an event signature is reused across distinct functions or execution paths with different semantics, where the emitted events have overlapping parameter domains. If off-chain systems only observe the event logs, the origin path is ambiguous and attackers can craft inputs that imitate another path, undermining event-based validation.

\begin{figure}[t]
    \small
\centering
{\scriptsize
\begin{lstlisting}
event Deposit(address indexed user, address indexed token, uint256 amount);
function depositETH() external payable {
    emit Deposit(msg.sender, address(0), msg.value);} // ETH is transferred via msg.value
function deposit(address token, uint256 amount) external {
    // missing: require(token != address(0)), returns false if token = zero
    token.transferFrom(msg.sender, address(this), amount);
    emit Deposit(msg.sender, token, amount);}
\end{lstlisting}
}
\caption{An example of Event Collision.}
\Description{Event collision code example. The depositETH function and the token deposit function emit indistinguishable Deposit events; the token path lacks a nonzero-token check.}
\label{fig:EC}
\end{figure}

\noindent{\bfseries Example:} \cref{fig:EC} shows a simple bridge contract that emits a \texttt{Deposit} event from both the \texttt{depositETH} and \texttt{deposit} functions (Line~3 vs. Line~7). If \texttt{deposit} does not enforce \texttt{token != address(0)} (Line~5), an attacker can invoke \texttt{deposit} with \texttt{token = address(0)} to emit a \texttt{Deposit} event that is indistinguishable from a genuine ETH deposit without transferring any ETH. Off-chain validators that rely on \texttt{Deposit} logs can accept it and release assets on the destination chain; this class of event collision underlies incidents such as Qubit Bridge~\cite{Qubit} and Meter Bridge~\cite{Meter}, with losses of approximately \$80 million and \$4.4 million, respectively.

\subsubsection{State-Event Mismatch (SEM)}
\revised{State--Event Mismatch occurs when an emitted event is inconsistent with the on-chain state transition it purports to summarize. For example, a contract may emit a \texttt{Transfer} event without the corresponding balance update, or log a value that differs from the updated state. Such mismatches violate the common off-chain assumption that events faithfully summarize state changes.}

\begin{figure}[t]
    \small
\centering
{\scriptsize
\begin{lstlisting}
event Transfer(address indexed from, address indexed to, uint256 value);
function transfer(address to, uint256 value) external {
    emit Transfer(msg.sender, to, value); } // missing balance updates
\end{lstlisting}
}
\caption{An example of State-Event Mismatch.}
\Description{State-event mismatch code example. A Solidity transfer function emits a Transfer event but performs no sender or receiver balance update.}
\label{fig:SEM}
\end{figure}

\noindent{\bfseries Example:} \cref{fig:SEM} shows a contract that emits a \texttt{Transfer} event (Line~3) while omitting the corresponding balance updates for the sender and receiver. Off-chain systems that rely on \texttt{Transfer} logs will record a transfer that never occurred on-chain.

\subsubsection{Unauthorized Event Emission (UEM)}
Unauthorized event emission refers to privileged events that untrusted callers can emit due to missing or insufficient access control, including cases where events are emitted before role or invariant checks. This misleads off-chain systems into inferring privileged actions that did not occur.

\begin{figure}[t]
    \small
\centering
{\scriptsize
\begin{lstlisting}
event LogEvent(address indexed _contract, address indexed _caller, bytes _data);
function Log(address _contract, address _caller, bytes memory _data) public {
    emit LogEvent(_contract, _caller, _data);} // no guard
\end{lstlisting}
}
\caption{An example of Unauthorized Event Emission from the DeFi Saver V3 audit.}
\Description{Unauthorized event emission code example. A public Log function emits caller-supplied event data without an authorization guard.}
\label{fig:UEM-real}
\end{figure}

\noindent{\bfseries Example:} The ConsenSys Diligence audit of DeFi Saver identified a missing access control check on the logging function \texttt{Log}; Line~3 shows an unrestricted emission. This can mislead off-chain monitors, as shown in \cref{fig:UEM-real}~\cite{Consensys-Mar-2021}.

\subsubsection{Event Emission Mismatch (EEM)}
\revised{Event Emission Mismatch concerns the event-emission behavior itself: the contract emits the wrong event type, omits an expected event, emits an unexpected or extra event, or emits the right event with an incorrect count or ordering. For example, a withdrawal function may emit \texttt{Deposit} instead of \texttt{Withdraw}, or a single logical action may produce duplicated events through multiple internal calls or alternative paths. These errors mislead off-chain observers about which actions actually occurred.}

\begin{figure}[t]
    \small
\centering
{\scriptsize
\begin{lstlisting}
event Deposit(address indexed user, uint256 amount);
event Withdraw(address indexed user, uint256 amount);
function withdraw(uint256 amount) external {
    _balances[msg.sender] -= amount;
    emit Deposit(msg.sender, amount);} // mismatch event
function deposit() payable external {
    _balances[msg.sender] += msg.value;}
\end{lstlisting}
}
\caption{An example of Event Emission Mismatch.}
\Description{Event emission mismatch code example. The withdraw function updates the balance but emits a Deposit event, while the deposit function updates the balance without emitting an event.}
\label{fig:EEM}
\end{figure}

\noindent{\bfseries Example:} As illustrated in \cref{fig:EEM}, a withdrawal function emits a \texttt{Deposit} event instead of \texttt{Withdraw} (Line~5), so observers see an unexpected event and miss the expected one, leading off-chain systems to record incorrect balances or actions.

\subsubsection{Event Parameter Mismatch (EPM)}
\revised{Event Parameter Mismatch occurs when an emitted event has the expected event type, but its parameters do not match the intended event schema or semantics. Examples include wrong parameter order (e.g., swapped \texttt{from}/\texttt{to} fields), incorrect indexed/non-indexed placement, or logging a stale or unrelated value. Such errors cause automated decoders and monitors to misinterpret the event.}

\begin{figure}[t]
    \small
\centering
{\scriptsize
\begin{lstlisting}
event Transfer(address indexed from, address indexed to, uint256 value);
function transfer(address to, uint256 value) external {
    _balances[msg.sender] -= value;
    _balances[to] += value;
    emit Transfer(to, msg.sender, value); // parameter order error
}
\end{lstlisting}
}
\caption{An example of Event Parameter Mismatch (parameter error).}
\Description{Event parameter mismatch code example. A Solidity transfer function emits a Transfer event with the intended sender and receiver parameters reversed.}
\label{fig:EPM}
\end{figure}

\noindent{\bfseries Example:} \cref{fig:EPM} shows a \texttt{Transfer} event where the \texttt{from} and \texttt{to} parameters are swapped (Line~5), resulting in an inverted transfer direction.

\subsection{Off-Chain Attack Vectors}
\label{sec:offchain-vectors}
We abstract two recurring off-chain attack vectors. These vectors capture how misleading event semantics propagate into off-chain systems, such as validation systems, indexing explorers, and user-facing services.
These vectors concern off-chain system behavior instead of contract-level defects, and emerge when event logs are accepted without validating the emitter or state.

\subsubsection{Event Origin Confusion Attack (EOCA)}
Event Origin Confusion occurs when off-chain systems accept event logs from an unintended emitter, typically due to weak emitter validation or overly broad/incorrect whitelisting, especially when a single transaction contains logs from both a legitimate emitter and a malicious one.

\noindent{\bfseries Example:} \cref{fig:eoca} illustrates this attack, and the pNetwork hack is a real-world case in which validators accepted events from an unintended contract, leading to unauthorized asset minting and theft of approximately \$13M~\cite{pNetwork}.

\subsubsection{Event-State Desynchronization Attack (ESDA)}
This attack arises when off-chain systems infer behavior solely from events without validating on-chain state or storage updates.

\begin{figure}[t]
    \small
    \centering
    \begin{subfigure}[b]{0.48\textwidth}
        \centering
        \includegraphics[width=\linewidth]{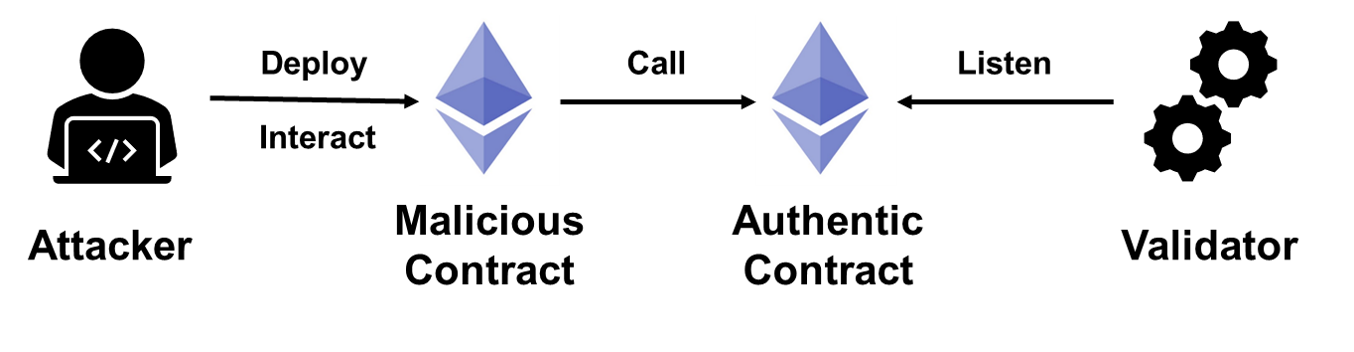}
        \caption{EOCA.}
        \label{fig:eoca}
    \end{subfigure}
    \hfill
    \begin{subfigure}[b]{0.48\textwidth}
        \centering
        \includegraphics[width=\linewidth]{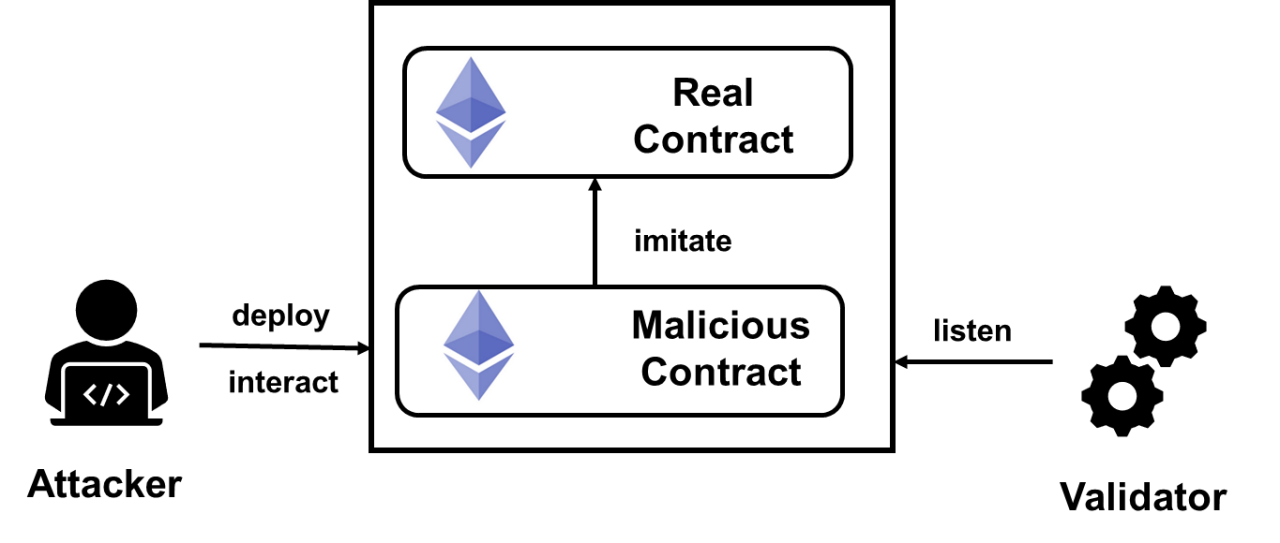}
        \caption{ESDA.}
        \label{fig:esda}
    \end{subfigure}
\caption{Two off-chain attack vectors.}
\Description{Two off-chain attack flows. In an event origin confusion attack, a malicious emitter produces a log that an off-chain consumer attributes to a legitimate contract. In an event-state desynchronization attack, a contract emits a transfer-like log without the corresponding state update, and the consumer records a nonexistent change.}
\label{fig:offchain-attacks}
\end{figure}

\noindent{\bfseries Example:} \cref{fig:esda} shows this pattern, and the sleepminting incident is a representative case where forged NFT transfer events, despite no real ownership change, were consumed by NFT marketplaces that relied exclusively on event logs~\cite{sleepminting}.

%% file: Sections/4_Methodology.tex
\section{Methodology}
\label{section:methodology}
\begin{figure*}[!t]
    \small
    \centering
    \includegraphics[width=\textwidth]{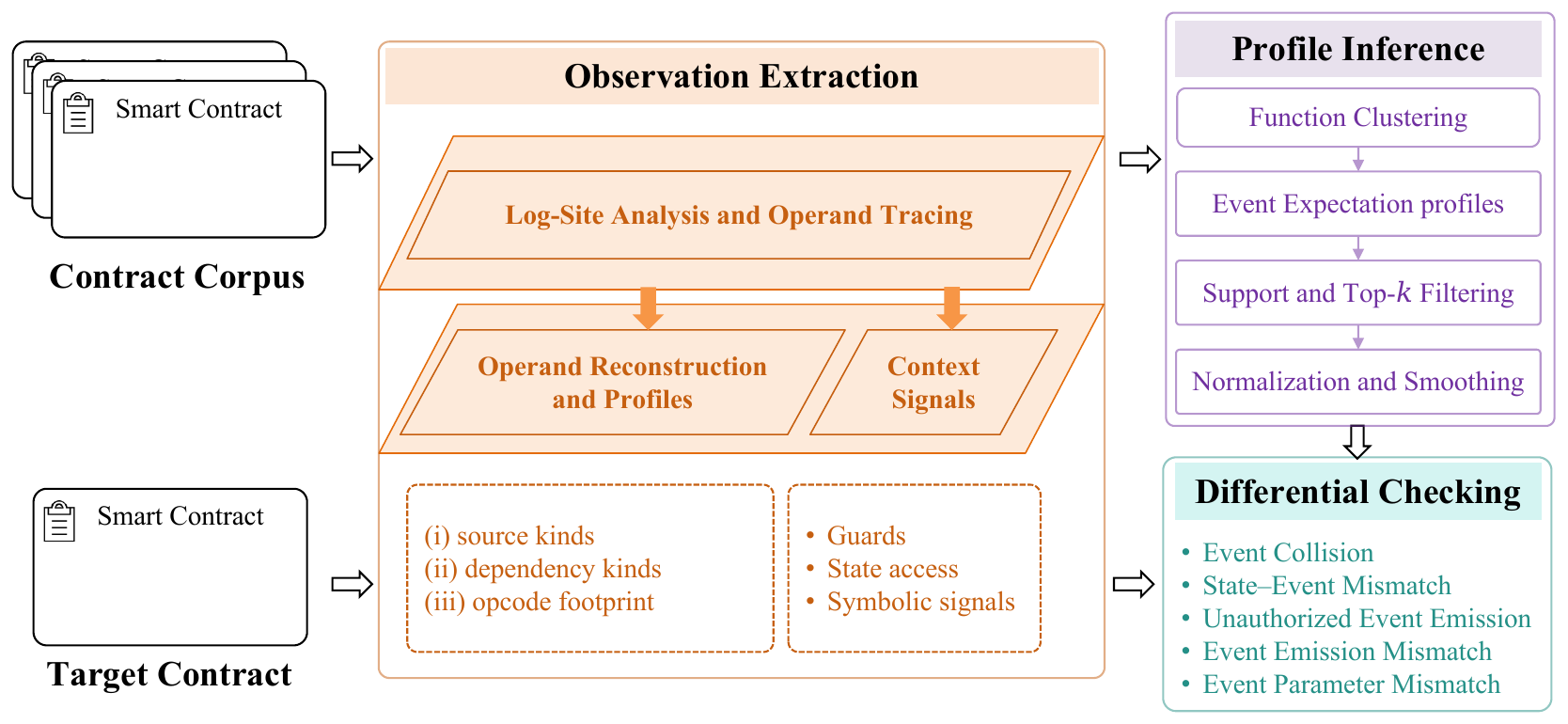}
    \caption{Workflow of \tool.}
    \Description{EventSpec workflow. Corpus bytecode is decompiled and analyzed to extract log-site observations, which are aggregated into event and function profiles. Target bytecode follows the same extraction path, and differential checking compares its observations with the inferred profiles to produce evidence-backed findings. A separate harness tests off-chain consumers.}
    \label{fig:methodology}
\end{figure*}
This section presents \tool, which constructs an empirical event specification from a contract corpus and performs differential checking on target contracts to detect event-semantic defects, complemented by an off-chain testing module. \tool is corpus-driven, bytecode-oriented, and produces explainable findings.

\subsection{Overview}

\tool takes two inputs: (i) a contract corpus to learn typical event-emission behavior, and (ii) a target contract to be analyzed. Each contract bytecode is decompiled into three-address code (TAC); per-function control-flow graphs (CFGs) are built over TAC basic blocks, and an interprocedural control-flow graph (ICFG) links functions for bounded guard attribution. State-write bindings are computed intra-procedurally. As shown in \Cref{fig:methodology}, the pipeline runs three modules: Extract captures observations at each log site, Infer aggregates them into event and function semantic profiles that form empirical event-emission specifications, and Diff compares target observations against these profiles. The output is a set of findings, each linked to a profile and supported by concrete evidence at the log and function levels.

\subsection{Notation and Schema}
We explicitly define an \emph{event specification} as the set of event and function level
profiles that summarize (i) layout information such as event signature hashes and indexed positions,
(ii) emission behavior such as per-function emission frequency for each event type and
log-shape distributions, and (iii) semantic constraints including parameter origins
(argument/state/literal), guard predicates, and state-write bindings expected at emission sites.
Each specification element is annotated with support and confidence statistics to enable
confidence-gated checking.

Let $\mathcal{F}$ denote the set of externally callable functions and $\mathcal{L}$ the set of log sites
in the analyzed contract. Each log site $\ell \in \mathcal{L}$ emits a \texttt{LOG} instruction with an
event signature hash $t_\ell$ (i.e., \texttt{topic0}) and a label set $L_\ell$ over topics and data.
Each emitted log is modeled as an \emph{event observation}
$o = \langle t, L, \Phi, G, W, B \rangle$, where $t$ is the event signature hash, $L$ is the label set over topics and data (e.g.,
\texttt{topic0..3}, \texttt{data0..n}), $\Phi$ captures per-label operand profiles, $G$ summarizes
access-control guards, $W$ summarizes log-reachable state writes, and $B$ captures operand--state
binding evidence between event operands and reachable storage reads or writes.

For each event type $t$, let $\pi(t)$ denote the matched event profile inferred from the corpus.
We treat $\pi(t)$ as an aggregated observation with the same fields $(L, \Phi, G, W, B)$ as a
log-level observation. During differential checking, each log site $\ell$ is compared against
its corresponding profile $\pi(t_\ell)$.

\subsection{Observation Extraction}

To instantiate event observations at each \texttt{LOG} site, we extract operand-level
data-dependency signals and guard evidence from TAC and CFG structures, and summarize log-reachable
state writes and bindings for the inference and differential checking modules.

\subsubsection{Log-Site Analysis and Operand Tracing}
\begin{figure}[!t]
    \small
    \centering
    \begin{subfigure}[b]{0.47\textwidth}
        \centering
        \includegraphics[height=2.55cm]{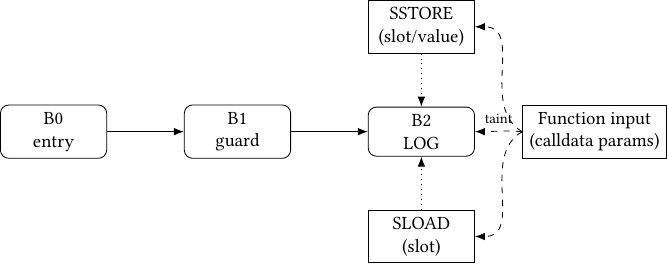}
        \caption{CFG slice around a log site.}
        \label{fig:cfg-slice}
    \end{subfigure}
    \hfill
    \begin{subfigure}[b]{0.47\textwidth}
        \centering
        \includegraphics[height=2.55cm]{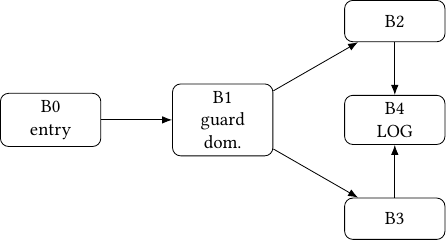}
        \caption{Dominator-based guard analysis.}
        \label{fig:dominator-guard}
    \end{subfigure}
    \caption{Log-site analysis: CFG slicing/operand tracing (left) and dominator-based guard inference (right).}
    \Description{Two control-flow diagrams. The left panel traces calldata, storage reads, and storage writes into an event-log instruction through a backward control-flow graph slice. The right panel shows an entry block leading to a dominating guard block, two branches, and a common event-log block, illustrating mandatory guard inference.}
    \label{fig:logsite-guard}
\end{figure}

To characterize each log site's operands and guards, we compute a log-reachable backward slice for
each operand and derive an opcode footprint (\Cref{algo:logslice}, \Cref{fig:cfg-slice}).
Separately, we identify dominator basic blocks for each log site as mandatory guard evidence used
later to derive guard signals (\Cref{fig:dominator-guard}).

\begin{algorithm}[!t]
\small
\footnotesize
	\caption{Log-Site CFG Slicing and Operand Tracing}
	\label{algo:logslice}

	\textbf{Input:} Bytecode \qquad
	\textbf{Output:} $\{\tau_{\ell,\lambda}\mid \ell \in \mathcal{L},\, \lambda \in \mathcal{A}_\ell\}$

\SetKwProg{Fn}{Procedure}{:}{}
\SetAlgoLined

IR $\gets$ Decompile(Bytecode)\;
Build CFGs and ICFG with reachability\tcp*{for log-reachable slicing}
$\mathcal{L} \gets \{\ell \mid \ell \text{ is a } \texttt{LOG} \text{ site}\}$;

	\ForEach{$\ell \in \mathcal{L}$}{
	  $\mathcal{A}_\ell \gets Operands(\ell)$\tcp*{topics \& data}

	  \ForEach{$\lambda \in \mathcal{A}_\ell$}{
	    $Slice_\lambda \gets BackSlice(\lambda,\mathrm{CFG})$\tcp*{log-reachable slice subgraph}
	    $T_\lambda \gets Vars(\lambda)$\tcp*{taint starts from the operand (sink)}
	    $T_\lambda \gets \textsc{TaintAnalysis}(Slice_\lambda, T_\lambda)$\tcp*{slice-based taint closure}
	    $\tau_{\ell,\lambda} \gets \{op(v) \mid v \in Slice_\lambda \land (\mathrm{DefVars}(v)\cup \mathrm{UseVars}(v)) \cap T_\lambda \neq \emptyset\}$\tcp*{tainted opcode footprint}
	  }
	}

\Fn{\textsc{TaintAnalysis}($S, T$)}{
  $T_{old} \gets \emptyset$\;
  \While{$T \neq T_{old}$}{
    $T_{old} \gets T$\;
    \ForEach{$v \in S$}{
      \If{$\mathrm{DefVars}(v) \cap T \neq \emptyset$}{
        $T \gets T \cup \mathrm{UseVars}(v)$\tcp*{backward taint step}
      }
    }
  }
  \KwRet $T$\;
}

	\textbf{return} $\{\tau_{\ell,\lambda}\mid \ell \in \mathcal{L},\, \lambda \in \mathcal{A}_\ell\}$\;
	\end{algorithm}

	\tool first decompiles smart contract bytecode into TAC annotated with opcodes, \texttt{DefVars} and
	\texttt{UseVars} sets over TAC variables, and basic-block identifiers.
	As shown in \Cref{algo:logslice}, operand-level opcode footprints are computed in three steps.
	First, \tool builds per-function CFGs and interprocedural control-flow graph (ICFG) connectivity, and computes reachability to each \texttt{LOG} site to restrict subsequent analysis to log-reachable statements (Lines~1–3).
	Second, for each \texttt{LOG} site $\ell$ and each of its operand labels $\lambda$, \tool constructs a log-reachable backward slice $Slice_\lambda$ over the CFG and initializes taint from the operand variables (Lines~4–9).
	Third, a slice-bounded backward taint analysis is applied over $Slice_\lambda$ to compute the taint closure $T_\lambda$, and the opcode footprint $\tau_{\ell,\lambda}$ is derived as the set of opcodes whose definitions or uses intersect $T_\lambda$ within the slice (Lines~10–25).
	The resulting opcode footprint $\tau_{\ell,\lambda}$ serves as an operand-level feature in $\Phi$ and is used by the profile inference and differential checking modules.

To conservatively capture operand origins under bounded analysis, we treat function-input sources from calldata (e.g., \texttt{CALLDATALOAD}/\texttt{ARG}), environment sources (e.g., \texttt{CALLER}, \texttt{CALLVALUE}, \texttt{ORIGIN}), and memory/storage reads (e.g., \texttt{MLOAD}, \texttt{SLOAD}) as taint sources when deriving operand source/dependency kinds. Dependencies are propagated along def-use chains within the log-reachable slice.

\Cref{fig:cfg-slice} illustrates the log-reachable slice subgraph produced by \Cref{algo:logslice}.
For each operand, we trace its data dependencies within this slice and summarize the derivation as
an opcode footprint, providing a lightweight signature without full path exploration.

\Cref{fig:dominator-guard} illustrates dominator-based guard inference. For a log site in basic
block $b_{\ell}$, any basic block $d$ that dominates $b_{\ell}$ must be executed on all paths that
reach the log site; therefore, predicates in $d$ form mandatory emission conditions (e.g., in the
figure, $B1$ dominates the log block). We extract guard evidence from such dominator blocks (e.g.,
\texttt{require} checks or branch conditions) and map them to guard categories, which are
also used by the inference and differential checking modules.

\subsubsection{Operand Reconstruction and Profiles}
Each log site $\ell$ has shape $s_\ell = (n_{t}, n_{d})$,where $n_{t}$ is the number of topics
	(indexed operands) and $n_{d}$ is the number of data words (non-indexed operands)..
Indexed operands can read from the topic region. In contrast, non-indexed operands are not directly accessible and must be reconstructed from the data region by locating relevant \texttt{MSTORE}
writes in the current basic block and its predecessors, resolving pointer arithmetic, and
aligning offsets to 32-byte slots. Each aligned slot is assigned a data label (\texttt{data0}, \texttt{data1}, \ldots) to represent reconstructed non-indexed operand. Auxiliary attributes such as literal class (\texttt{zero}/\texttt{nonzero}) and argument index (when operand originates from calldata params, i.e., \texttt{ARG}/\texttt{CALLDATALOAD}) are derived to refine operand profiles during matching and comparison. For each operand label $\lambda$, operand profile includes:
(\emph{i}) source kinds
(\emph{ii}) dependency kinds (e.g., \texttt{arith}, \texttt{bitwise},
\texttt{compare}, \texttt{hash}, \texttt{mem}, \texttt{storage}, \texttt{data},
\texttt{env}, \texttt{call}) and
(\emph{iii}) opcode footprint from the def-use path. \textit{Lightweight constraint} from the def-use path, including comparisons against constants, nonzero
checks, masking or shifts.

\subsubsection{Context Signals}
Contextual signals are captured at each log site and explicitly mapped to guard validation, state–event consistency checking, and event collision detection. For \textbf{guards}, we classify categories by opcode-structure
heuristics over dominator blocks, aligned with common permission-constraint
patterns summarized by PrettySmart~\cite{zhong2024prettysmart}, including owner checks,
role-based checks, whitelist lookups, and data-driven checks; when an internal
function containing a log site lacks local guards, caller functions are searched
via a bounded interprocedural traversal, and guard categories are attributed from
the call chain, since internal helpers are not directly callable and their
effective guards are enforced at external entry points in $\mathcal{F}$. For
\textbf{state updates and reads}, we restrict \texttt{SSTORE} statements to those
in the same function and on CFG paths that reach the basic block containing the log site with statement
order before the log, summarize slot classes (constant / linear / mapping-like)
inferred from slot-expression forms, record two binding modes for writes
(\emph{value-binding} when an operand path intersects the \texttt{SSTORE} value
path, and \emph{slot-binding} when it intersects the \texttt{SSTORE} slot path),
and record storage-read bindings when an operand path intersects the slot path
of a reachable \texttt{SLOAD}; these bindings are used to flag state-event
mismatches, including inputs emitted directly from calldata parameters
(\texttt{ARG}/\texttt{CALLDATA}) without any binding to reachable storage
reads/writes. For \textbf{signatures and symbolic signals}, $t_0$ is matched to
the known event signature when available and the hash remains the primary key for
matching and inference; path constraints and symbolic expressions are extracted
from the taint slice and guard predicates, and SMT solving is applied at
comparison to assign each operand pair an equality status (\texttt{possible} for
satisfiable, \texttt{unsat} for provably unequal), with unknown results
discarded.

\subsection{Profile Inference}
\subsubsection{Function Clustering}
Functions are clustered by a behavioral signature. We define
$\mathrm{sig}(f)=\mathrm{hash}(G_f, S_f, D_f, M_f, O_f)$, where $G_f$ is the
guard-category set, $S_f$ the storage-write kinds, $D_f$ the frequent operand
dependency/source kinds, $M_f$ the modal (topic count, data count) pair across
log sites, and $O_f$ the frequent opcodes. Functions with the same signature
form a cluster $c$.

\subsubsection{Event Expectations and Shape Profiles}
	At the cluster level, the expected probability of emitting a given event type is
	estimated as $P_c(t_0)=n_c(t_0)/|F_c|$, where $n_c(t_0)$ counts functions in
	cluster $c$ that emit $t_0$ and $F_c$ is the set of functions in $c$. The distribution of total log counts and
	per-topic log counts, which enables checks for missing or over-emitted events,
	is included. For each $t_0$, we use a log-shape distribution over $(n_t,n_d)$, where $n_t$ and $n_d$
	denote the topic and data label counts,
	together with the indexed-label set (e.g., \texttt{topic1..}) to capture
	indexed/data layout expectations; deviations from the dominant shape are treated
	as potential parameter mismatches. Other per-label operand, guard, and binding
	profiles are aggregated from counts and used in similarity scoring.

\subsubsection{Support and Top-$k$ Filtering}
Support denotes the number of contributing functions for function-semantic profiles and
the number of event observations for event-semantic profiles (per cluster and $t_0$).
To control noise for common events, only the top-$k$ event-semantic profiles per $t_0$
by support are retained, and differential checking enforces a minimum support
threshold.

\subsubsection{Normalization and Smoothing}
\label{sec:smooth}
Before differential checking, frequency profiles are normalized into probability
distributions. For distributional similarities we add an $\alpha$ pseudo-count
to each label before normalization, and for Bernoulli-style presence checks we
clamp probabilities using the same $\alpha$ to avoid 0/1 extremes~\cite{chen1999smoothing}. This makes similarity scores stable even when a
profile has limited support and prevents a single missing label from collapsing
the overall score.

\subsection{Differential Checking}
Differential checking compares a target contract against the corpus-derived
specification. We match extracted events and functions to their closest inferred
profiles, score similarity over the corresponding feature sets, and apply
confidence gating (minimum similarity, minimum support, and Wilson-interval
constraints) before recording taxonomy-aligned findings with evidence. In
parallel, we run target-only checks that do not depend on profile matching,
including missing-log detection, event collision checks, and insufficient-guard
checks.

\subsubsection{Profile Matching and Similarity}
For a target contract, observations are extracted and each event observation is matched to
the best event profile for its $t_0$, and each function is matched to the best
function profile. Event similarity is a weighted combination of indexed-label
likelihood, parameter-count likelihood, operand source similarity, operand
dependency similarity, \texttt{SSTORE}-kind overlap, and guard-category overlap:
{\small
\[
\mathrm{sim}_e =
\frac{\sum_{k} w_k\,\mathrm{sim}_k}{\sum_{k} w_k}
\]
}%
Indexed-label similarity uses the average log-likelihood of label presence under
the indexed profile, and count similarity uses the probability of the observed
topic/data counts. Operand source/dependency similarities use
Jensen-Shannon similarity~\citep{lin1991divergence} over per-label distributions,
and guard/SSTORE overlap is measured by Jaccard similarity.
Function similarity is computed as $\mathrm{sim}_f = w'_{st}\,\mathrm{sim}_{st} + w'_{gd}\,\mathrm{sim}_{gd} + w'_{op}\,\mathrm{sim}_{op}$, where $\mathrm{sim}_{st}$ and $\mathrm{sim}_{gd}$ are Jaccard overlaps over storage-write kinds and guard categories, and $\mathrm{sim}_{op}$ is the Jaccard overlap of high-frequency opcodes; the weights $w'_*$ sum to 1.

\subsubsection{Confidence Gating and Thresholds}
Findings are emitted only when similarity exceeds a minimum matching threshold and support exceeds
a minimum support count. To avoid brittle decisions on small samples, each probability-based decision is gated with a Wilson
confidence~\citep{wilson1927probable}. For an estimated probability $p$ computed from $n$
supporting samples, the interval width is:
{\small
\[
w = \frac{2z}{1 + z^2/n} \sqrt{\frac{p(1-p)}{n} + \frac{z^2}{4n^2}}
\]
}
We require $w \leq w_{\max}$, where $z$ is the normal critical value for the chosen confidence level
and $w_{\max}$ is the maximum allowed width. Guard and storage checks have their own thresholds, and
operand-profile checks use label-specific thresholds for source/literal/argument index. All probability
computations here use the smoothed distributions described in~\cref{sec:smooth}.

\subsubsection{Findings}

\begin{figure}[!t]
		    \small
		    \centering
		    \begin{tabular}{@{}p{0.18\linewidth}p{0.36\linewidth}p{0.36\linewidth}@{}}
		        \toprule
		        \textbf{Field group} & \textbf{Profile $\pi(t_\ell)$} & \textbf{Target $o$ at log site $\ell$} \\
		        \midrule
			        \textbf{Layout} &
			        \makecell[l]{\footnotesize
			            $t_\ell$\\
			            $(n_t,n_d)=(3,1)$\\
			            $L=\{\mathrm{topic0..2},\mathrm{data0}\}$
			        } &
		        \makecell[l]{\footnotesize
		            $t_\ell$\\
		            $(n_t,n_d)=(3,1)$\\
		            $L=\{\mathrm{topic0..2},\mathrm{data0}\}$
		        } \\
	        \cmidrule(lr){1-3}
	        \textbf{Operands} &
	        \makecell[l]{\footnotesize
	            $\Phi$: src(topic1)=CONST\\
	            footprint: \dots
	        } &
	        \makecell[l]{\footnotesize
	            $\Phi$: src(topic1)=CONST\\
	            footprint: \dots
	        } \\
	        \cmidrule(lr){1-3}
		        \textbf{Guards} &
		        \makecell[l]{\footnotesize
		            $G$: owner (dominant)
		        } &
		        \makecell[l]{\footnotesize
		            \textcolor{red}{$G$: none}
		        } \\
	        \cmidrule(lr){1-3}
	        \textbf{Constraints} &
	        \makecell[l]{\footnotesize
	            $B$: data0 $\leftrightarrow$ \texttt{SSTORE.value}\\
	            \dots
	        } &
	        \makecell[l]{\footnotesize
	            $B$: \dots
	        } \\
	        \bottomrule
	    \end{tabular}
    \caption{Example: missing guard evidence.}
    \Description{Missing guard evidence table. The table compares an inferred event profile with a target log observation in three columns. Layout and operand fields match, while the profile requires a dominant owner guard and the target has no guard; the target also lacks the profile's state-write binding.}
	    \label{fig:finding-example}
	\end{figure}

The Diff module produces findings aligned with the taxonomy in \Cref{table:Definition}.
Each finding is instantiated by comparing a target observation against its matched profile
and applying a category-specific detection rule that combines profile-derived expectations
with target-side consistency checks, including SMT-backed signals.

\Cref{fig:finding-example} illustrates the evidence used by the differential checker.
Each target log site is compared against its matched event profile $\pi(t_\ell)$, and each
target function is compared against its matched function profile to derive expected
emission behavior. The example highlights missing guard evidence, which
triggers an Unauthorized Event Emission finding when the log-emitting function has an
empty guard summary.

For an observation $x$, we denote its label set, log shape, and operand profiles as
$L_x$, $s_x$, and $\Phi_x$, respectively. We use $\mathrm{Eq}(\cdot)$ to denote the
conjunction of SMT-encodable operand-equality constraints and $\mathrm{SAT}(\cdot)$ to
denote satisfiable constraints. For binding evidence, $\mathrm{Bound}_{x}(\lambda)$
indicates that label $\lambda$ has any operand--state binding recorded via log-reachable
storage reads or writes. 

We classify findings by applying the detection rules summarized in
\Cref{table:diff-rules}. For function-level comparisons, we write
$\mathrm{Count}_{f}(t_0)$ for the expected emission count of event type $t_0$ under the
matched function profile of a target function $f$, subject to the confidence and
thresholding constraints. In brief, ``Event Collision''
flags indistinguishable emissions whose operand constraints overlap; ``State--Event
Mismatch'' flags disagreement between expected and observed operand--state binding
evidence; ``Unauthorized Event Emission'' flags log-emitting functions lacking any guard
evidence after interprocedural attribution; ``Event Emission Mismatch'' flags deviations
between expected and observed per-function event counts; and ``Event Parameter Mismatch''
flags inconsistencies in log layout or per-label operand profiles relative to the matched
event profile.
		
\begin{table*}[t]
\centering
\footnotesize
\caption{Detection rules for differential finding categories.}
\Description{Differential detection rules. The table maps each event-semantic finding category to the logical condition used to detect it from inferred and observed event profiles.}
\label{table:diff-rules}
\begin{tabular}{@{}p{0.29\textwidth}@{\hspace{4pt}\vrule\hspace{4pt}}p{0.67\textwidth}@{}}
\toprule
\textbf{Finding} & \textbf{Detection rule} \\
\midrule

\textbf{\mbox{Event Collision}} &
\parbox[t]{\linewidth}{$\textstyle
\exists \ell_1\neq \ell_2:\ 
t_{\ell_1}=t_{\ell_2}\ \land\ 
s_{\ell_1}=s_{\ell_2}\ \land\ 
L_{\ell_1}=L_{\ell_2}\ \land\ 
\mathrm{SAT}(\mathrm{Eq}(o_{\ell_1},o_{\ell_2})) 
$} \\
\cmidrule{1-2}

\textbf{\mbox{State--Event Mismatch}} &
\parbox[t]{\linewidth}{$\textstyle
\exists \ell\in \mathcal{L},\exists \lambda\in (L_\ell\cap L_{\pi(t_\ell)}):\
\mathrm{Bound}_{\ell}(\lambda)\neq \mathrm{Bound}_{\pi(t_\ell)}(\lambda)
$} \\
\cmidrule{1-2}

\textbf{\mbox{Unauthorized Event Emission}} &
\parbox[t]{\linewidth}{$\textstyle
\exists f\in \mathcal{F},\ \exists \ell\in \mathcal{L}:\ 
\ell\in f\ \land\ 
G_f=\emptyset \ \land\ G_{\pi(f)}\neq \emptyset
$} \\
\cmidrule{1-2}

\textbf{\mbox{Event Emission Mismatch}} &
\parbox[t]{\linewidth}{$\textstyle
\exists f\in \mathcal{F},\exists t_0:\ 
\mathrm{Count}_{f}(t_0)\neq 
|\{\ell\in \mathcal{L}\mid \ell\in f\land t_\ell=t_0\}|
$} \\
\cmidrule{1-2}

\textbf{\mbox{Event Parameter Mismatch}} &
\parbox[t]{\linewidth}{$\textstyle
\exists \ell\in \mathcal{L}:\ 
(s_\ell\neq s_{\pi(t_\ell)}\ \lor\ L_\ell\neq L_{\pi(t_\ell)})\ \lor\
\exists \lambda\in (L_\ell\cap L_{\pi(t_\ell)}):\
\Phi_\ell[\lambda]\neq \Phi_{\pi(t_\ell)}[\lambda]
$} \\
\bottomrule
\end{tabular}
\end{table*}

\subsection{Off-Chain Evaluation Harness}
\label{sec:harness}
\revised{The harness is an \emph{evaluation} component, not part of the static analysis pipeline. Its purpose is to test whether off-chain consumers (wallets, explorers, bridge relayers) accept misleading event semantics that an on-chain contract could legitimately emit. We design two attacker-side test contracts.

\noindent\textbf{EOCA test contract.} In a single transaction, the contract calls the legitimate target so that a genuine event is emitted from the target's address, and emits a forged event of the same type from a separate attacker-controlled emitter address. An off-chain consumer that allowlists by event type alone, without binding each event to its intended emitter, will accept the forged event as genuine.

\noindent\textbf{ESDA test contract.} The contract emits an event (e.g., \texttt{Transfer}) from an attacker-controlled address \emph{without} performing the corresponding state update (e.g., balance change). An off-chain consumer that infers state purely from event logs, without cross-checking the on-chain state, will record a state change that never happened (the sleepminting pattern).

}

%% file: Sections/5_Evaluation.tex
\section{Evaluation}
\label{sec:evaluation}

\subsection{Research Questions}
In this section, we aim to answer the following research questions.
\begin{itemize}
  \item \textbf{RQ1:} How ERC-consistent are corpus-inferred event-emission
  specifications?
  \item \textbf{RQ2:} How accurate is \tool and how common are event-semantic issues in the wild?
  \item \textbf{RQ3:} How feasible is it to perform the two off-chain attack vectors in real-world systems?
\end{itemize}

\noindent\textbf{Implementation.} \tool uses Gigahorse~\cite{Gigahorse,lagouvardos2025shrnkr} to decompile bytecode into
TAC and construct CFGs, and symbolic equality checks are supported by the Greed
module~\cite{ruaro2025history}. The remaining components (extraction, inference, and differential
checking) are implemented in about 7k lines of Python, and the harness includes 51 lines of
Solidity.

\noindent\textbf{Configuration.} We evaluate \tool in a controlled environment on a workstation with 4 physical CPU cores (8
threads), \SI{64}{\giga\byte} RAM, and Ubuntu 22.04.5 LTS. We deduplicate contracts by bytecode hash. Default settings: smoothing $\alpha{=}1.0$, min support 20,
top-$k$ event profiles 3, event similarity threshold 0.65, function similarity threshold 0.65,
guard/binding threshold 0.8, confidence level $z{=}1.96$, and max interprocedural depth 6. We also
set a per-contract analysis time limit of \SI{1200}{\second}, a path exploration depth limit of 500, and an
SMT solver timeout of \SI{300}{\second}.

\subsection{Datasets}
\noindent\textbf{Data source.} We derive our datasets from the Smart Contract Sanctuary collection
of verified Etherscan contracts~\cite{smartcontractsanctuary}. We use the snapshot that includes
contracts verified up to July~13,~2023, and selected contract files
deployed on the Ethereum mainnet, totaling 331{,}382 mainnet contracts.

\noindent\textbf{Corpus (behavior inference).} We deduplicate
contracts by deployed bytecode and exclude contracts that exceed the \SI{150}{\second} decompilation
timeout, yielding 212{,}072 contracts.

\noindent\textbf{Target (detection pool).} We use \TargetPoolSize verified Ethereum smart contracts as the target pool for large-scale detection (RQ2), drawn from the NumScout-curated set~\cite{chen2025numscout} over Smart Contract Sanctuary~\cite{smartcontractsanctuary}. The set retains contracts with more than 100 historical transactions, a non-zero ether balance, and successful compilation under the recorded Solidity version, focusing the evaluation on mainstream deployed Ethereum smart contracts that are substantially larger and more modern than common benchmarks such as SmartBugs~\cite{ferreira2020smartbugs} (e.g., 11.5$\times$ LOC, 5.5$\times$ instructions, and a larger fraction compiled with Solidity~$\geq$~0.8.0). The 95-contract manually labeled dataset described below is a random subset of this target pool, not an additional pool.

\noindent\textbf{ERC standards dataset.} For ERC events, we derive event specifications directly from
ERC documents listed on the EIPs site~\cite{eips}. We include all ERC standards published on the
Ethereum Improvement Proposals site up to Jan.~20,~2026 with Final status. To focus on adopted standards, we map ERCs to implementations using the Smart Contract
Sanctuary dataset and exclude ERCs with fewer than 50 contracts to ensure sufficient support,
yielding 14 ERCs and 26 event definitions~\cite{smartcontractsanctuary}. We extract event
declarations to obtain layout, and annotate normative statements
to derive semantic constraints on parameter origins and required guards. Two authors
annotate normative statements and resolve disagreements through discussion.
 
\noindent\textbf{Manually labeled dataset.} We manually label contracts randomly sampled from the filtered dataset for the presence of the five issue types. \revised{Following the sampling protocol used in concurrent peer-reviewed work~\cite{chen2025numscout,SSR,tonscanner}, applying Cochran's formula with finite-population correction at 95\% confidence and a $\pm10\%$ margin of error gives $n = 95$ for the pool of 6{,}617 contracts.} Two authors jointly label 30\% of the samples to establish labeling criteria, then independently label the remaining 70\% and resolve disagreements through discussion. \revised{Inter-rater agreement on the independently labelled 70\% subset was $\kappa = 0.97$ (\emph{almost perfect}), and any remaining labelling disputes were arbitrated by two additional authors.} \revised{This labeled subset serves as the ground truth for the per-category detection accuracy reported in RQ2.}

\begin{table}[t]
  \centering
  \small
  \caption{RQ1 sensitivity to inference parameters.}
  \Description{Inference-parameter sensitivity. Rows vary smoothing, minimum support, and the number of retained profiles, and report the number of profiles and their signature, layout, and semantic match rates.}
  \label{tab:rq1-sensitivity}
  \begingroup
  \setlength{\tabcolsep}{1.6pt}
  \begin{tabular}{l c c c c c}
    \toprule
    \textbf{Parameter} & \textbf{Value} & \textbf{Profiles} & \textbf{Signature match} & \textbf{Layout match} & \textbf{Semantic match} \\
    \midrule
    \multirow{5}{*}{$\alpha$} & 0.1 & 1246 & 80.77\% & 100.00\% & 90.48\% \\
    & 0.5 & 1246 & 80.77\% & 100.00\% & 90.48\% \\
    & 1 & 1246 & 80.77\% & 100.00\% & 90.48\% \\
    & 2 & 1246 & 80.77\% & 100.00\% & 90.48\% \\
    & 5 & 1246 & 80.77\% & 100.00\% & 90.48\% \\
    \midrule
    \multirow{3}{*}{Min support} & 5 & 8510 & 100.0\% & 96.15\% & 88.46\% \\
    & 10 & 2954 & 96.15\% & 96.00\% & 88.00\% \\
    & 20 & 1246 & 80.77\% & 100.00\% & 90.48\% \\
    \midrule
    \multirow{5}{*}{Top-$k$} & 1 & 793 & 80.77\% & 80.95\% & 71.43\% \\
    & 3 & 1246 & 80.77\% & 100.00\% & 90.48\% \\
    & 5 & 1439 & 80.77\% & 100.00\% & 90.48\% \\
    & 10 & 1695 & 80.77\% & 100.00\% & 95.24\% \\
    \bottomrule
  \end{tabular}
  \endgroup
\end{table}

\begin{table}[t]
  \centering
  \small
  \caption{RQ1 sensitivity to the ERC inclusion threshold.}
  \Description{Token-standard threshold sensitivity. Rows vary the minimum number of contracts required to include a standard and report the retained event count and signature, layout, and semantic match rates.}
  \label{tab:rq1-erc-threshold}
  \begingroup
  \setlength{\tabcolsep}{2.0pt}
  \begin{tabular}{l c c c c}
    \toprule
    \textbf{Min contracts} & \textbf{Events} & \textbf{Signature match} & \textbf{Layout match} & \textbf{Semantic match} \\
    \midrule
    50 & 26 & 80.77\% & 100.00\% & 90.48\% \\
    100 & 22 & 95.45\% & 100.00\% & 90.48\% \\
    250 & 15 & 100.0\% & 100.00\% & 86.67\% \\
    1000 & 12 & 100.0\% & 100.00\% & 83.33\% \\
    \bottomrule
  \end{tabular}
  \endgroup
\end{table}
\subsection{RQ1: ERC Consistency of Inferred Specifications}
\revised{RQ1 provides a controlled validation setting for our corpus-driven inference engine: it evaluates whether specifications inferred from real-world implementations recover ERC event declarations, the only authoritative reference for Ethereum events. We therefore compare inferred specifications against ERC-declared layouts and normative semantic constraints before applying empirical inference to events that lack formal specifications.}
We evaluate corpus-inferred event-emission specifications for layout and semantic consistency
against ERC event declarations and normative statements.
We decompose the specifications into two components: (i) layout and (ii) a semantic profile.
Layout is defined as the event signature and indexed parameter positions, and the semantic profile
is defined by parameter origin constraints (argument-tainted, state-tainted, or unconstrained) and
required guard/binding conditions. A layout
match requires the correct signature in \texttt{topic0} and indexed positions. A semantic match requires that the inferred origin and guard/binding signals
satisfy all documented constraints.
We report layout/semantic match rates for ERC events with sufficient support for inference. For
ERC events with sparse support, we do not attempt to infer or evaluate semantics due to
insufficient evidence.

\noindent\textbf{Match rate definitions.}
Let $E$ be the set of evaluated ERC events and $E_{\text{sig}}\subseteq E$
the subset whose inferred \texttt{topic0} matches the ERC-declared
signature. The signature match rate is $|E_{\text{sig}}|/|E|$.
Layout and semantic match rates are computed over $E_{\text{sig}}$, as
$|E_{\text{layout}}|/|E_{\text{sig}}|$ and
$|E_{\text{semantic}}|/|E_{\text{sig}}|$, respectively, where
$E_{\text{layout}}$ and $E_{\text{semantic}}$ additionally satisfy
indexed-position and semantic constraints.

\noindent\textbf{RQ1 summary and parameter effects.}
On the ERC standards dataset,~\Cref{tab:rq1-sensitivity} shows that varying $\alpha$ does not
materially change layout or semantic matching. This is because $\alpha$ only smooths empirical
frequencies, and for ERC events with enough observations the dominant layout and origin/guard
signals already exceed decision thresholds. The same table also shows that signature non-matches
under the default setting are mainly due to limited ERC adoption after bytecode deduplication:
some ERC standards have few remaining implementations and fail the support threshold, while
lowering support allows these signatures to be inferred. With top-$k{=}1$, layout match rates
drop, indicating that the most common real-world usage of an event is not always the
ERC-required one, and ERC-compliant behavior can appear as a lower-support mode that is pruned
when only the single dominant pattern is kept.

\noindent\textbf{Implications and configuration choice.}
\Cref{tab:rq1-erc-threshold} shows that excluding low-adoption ERCs improves signature and layout
agreement, but shows no consistent improvement in semantic agreement, which can decrease as the
threshold increases. This indicates that the main limitation lies in semantic inference rather
than data sparsity. In particular, many ERC documents specify authorization or state-update
conditions that do not align with the limited guard types we currently infer, which also differ
from the five guard categories studied in PrettySmart~\cite{zhong2024prettysmart}. Based on these
observations, we use the default setting for RQ2 ($\alpha{=}1$, minimum support $=20$,
top-$k{=}3$), which gives stable inference, avoids low-support noise, and preserves
ERC-compliant patterns that are not the single most frequent mode. \revised{Taken together,
the high layout-level agreement supports the inference methodology, while semantic-level
disagreements show that real-world implementation patterns do not always conform to ERC
specifications, motivating empirical inference for events without formal specifications.}

\subsection{RQ2: Detection Accuracy and Prevalence}
\label{sec:rq2}

We estimate real-world prevalence over the complete target pool of \TargetPoolSize contracts, counting each issue type at most once per contract. Detection effectiveness is evaluated on the 95 manually labeled contracts from the same pool; five additional Event Collision cases collected during open card sorting are used only for Event Collision effectiveness.

We report precision/recall/F1 per issue type and an overall comprehensive precision. Metrics are computed at the
contract level: a contract is positive for an issue if any log site exhibits it, and a detection
is correct if \tool reports at least one matching finding for that issue in the contract.
Precision/recall/F1 follow standard definitions. We compute
comprehensive precision as $P_{\text{comp}}=\frac{\sum_i P_i \cdot |C_i|}{\sum_i |C_i|}$, where
$|C_i|{=}TP_i{+}FN_i$, and report macro-averaged recall across issue types. 
\begin{table*}[t]
\centering
\small
  \caption{RQ2 results: large-scale detection yield on the target pool of \TargetPoolSize contracts, and detection effectiveness on 95 labeled contracts (Event Collision uses five open card sorting contracts).}
  \Description{Event-semantic detection results. For each of five issue types, the table reports large-scale finding counts and percentages together with true positives, false positives, false negatives, precision, recall, and the harmonic-mean score on the labeled evaluation set.}
  \label{tab:rq2}
  \setlength{\tabcolsep}{2.2pt}
  \resizebox{\linewidth}{!}{%
  \begin{tabular}{c|c|c|c|c|c|c|c|c}
    \hline
    \textbf{Issue} & \textbf{\#} & \textbf{Pct.(\%)} & \textbf{TP} & \textbf{FP} & \textbf{FN} & \textbf{Precision(\%)} & \textbf{Recall(\%)} & \textbf{F1-score(\%)} \\
    \hline
    \hline
    \textit{Event Collision} & 0 & 0 & 5 & 0 & 0 & 100.00 & 100.00 & 100.00 \\
    \hline
    \textit{State-Event Mismatch} & 4159 & 62.85 & 45 & 9 & 0 & 83.33 & 100.00 & 90.91 \\
    \hline
    \textit{Unauthorized Event Emission} & 405 & 6.12 & 11 & 2 & 1 & 84.62 & 91.67 & 88.00 \\
    \hline
    \textit{Event Emission Mismatch} & 6514 & 98.44 & 90 & 1 & 1 & 98.90 & 98.90 & 98.90 \\
    \hline
    \textit{Event Parameter Mismatch} & 3664 & 55.37 & 18 & 9 & 2 & 66.67 & 90.00 & 76.60 \\
    \hline
  \end{tabular}%
  }
\end{table*}

\Cref{tab:rq2} summarizes RQ2 results, reporting per-issue metrics as well as the
overall comprehensive precision. Overall, the comprehensive precision reaches 90.17\%, and the
macro-average recall is 96.11\%.
State-Event Mismatch is frequently caused by smart contracts that emit events before
updating state, which violates common best practices; we treat such cases as true
positives. False positives for State-Event Mismatch and Event Parameter Mismatch mainly
arise when the inferred semantic profiles of smart contracts differ across event,
parameter, or state dimensions, despite representing the same underlying intent.
\revised{For EPM specifically, the lower precision mainly stems from semantic ambiguity at the parameter level: equivalent event semantics can be implemented using different parameter representations (e.g., derived values such as scaled storage reads, reordered fields, or indirect state references through helper functions), and our inference matches syntactic patterns rather than semantic equivalence.}
For Unauthorized Event Emission, false positives mainly stem from incomplete permission
inference, while false negatives are primarily caused by decompilation failures
that prevent accurate recovery of function blocks. Event Emission Mismatch errors are
largely due to limited inter-contract call analysis: although the current contract does
not emit the expected event, the emission occurs elsewhere along the call chain. In
general, issues associated with explicit guards or binding signals (e.g., Unauthorized
Event Emission and State-Event Mismatch) tend to be easier to detect, whereas issues
that rely on parameter-origin inference (e.g., Event Parameter Mismatch) remain more
challenging.

Overall, at least 6{,}514 out of \TargetPoolSize contracts (98.44\%) contain at least one tool-reported issue,
since Event Emission Mismatch alone appears in 6{,}514 contracts. The most prevalent issue type
is Event Emission Mismatch (98.44\%), followed by State-Event Mismatch (62.85\%) and Event
Parameter Mismatch (55.37\%). Unauthorized Event Emission appears in 6.12\% of contracts. Although Event Collision is rare in our datasets, it has led to real-world attacks and measurable losses in cross-chain applications. The high EEM yield is dominated by \texttt{no\_event\_emission}, \tool's operational rule for non-view functions with no reachable log but with state changes or other non-view operations. Thus, 98.44\% denotes findings under the detector definition rather than independently confirmed exploitable vulnerabilities in 98.44\% of contracts.

\revised{
\paragraph{Comparison with General-Purpose Smart Contract Analyzers.}
\label{sec:rq4-baselines}
To assess whether general-purpose smart-contract analyzers cover event-semantic defects, we run four widely used tools on the same 95 labelled contracts, spanning complementary analysis paradigms: Solidity-AST data-flow analysis (Slither~\cite{Slither}), industry-oriented SWC-based symbolic execution (Mythril~\cite{mythril}), XPath/rule-based pattern matching (SmartCheck~\cite{tikhomirov2018smartcheck}), and academic symbolic execution (Oyente~\cite{luu2016making}). We conservatively map each tool's documented detectors to our five categories, retaining a detector only when its documented semantics directly match one of our categories; ambiguous or non-event detectors are excluded rather than force-classified.

Under this mapping, only three Slither detectors qualify: \texttt{events-maths} and \texttt{events-access} detect missing events after arithmetic or access-control changes, covering only a subset of EEM, while \texttt{reentrancy-events} detects reentrancy-induced event-state desynchronization, covering a subset of SEM. Mythril, SmartCheck, and Oyente do not provide detectors designed for event-emission paths and therefore achieve 0\% F1 across all five categories. Accordingly, \Cref{tab:rq2-baselines} reports per-category F1 for \tool and Slither, the only baseline with non-zero F1. For Slither's two non-zero categories, SEM has low precision (17.74\%) and recall (24.44\%), while EEM has perfect precision (100.00\%) but limited recall (52.75\%). The low SEM precision is consistent with prior evidence that smart-contract SAST tools, including detectors such as Slither's \texttt{reentrancy-events}, exhibit high false-positive rates~\cite{li2024sast}.

Slither's coverage is also limited to narrow SEM/EEM sub-cases: it misses wrong event types, parameter swaps, event-signature collisions, and missing emissions unrelated to arithmetic or access-control changes. These results indicate that event-semantic defect detection remains a design gap in general-purpose smart-contract analyzers, rather than merely a matter of detector tuning.

\begin{table}[t]
\centering\small
\caption{Tool comparison: per-category F1 (\%) on the 95 labelled contracts.}
\Description{Analyzer comparison. The table reports per-category harmonic-mean scores for EventSpec and Slither across the five event-semantic issue types, plus the average across all categories.}
\label{tab:rq2-baselines}
\setlength\tabcolsep{4pt}
\begin{tabular}{l|ccccc|c}
\toprule
Tool & EC & SEM & UEM & EEM & EPM & Avg \\
\midrule
\textbf{\tool}            & 100.00 & 90.91 & 88.00 & 98.90 & 76.60 & 90.88 \\
Slither~\cite{Slither}    & --     & 20.55 & --    & 69.07 & --    & 17.92 \\
\bottomrule
\end{tabular}
\end{table}
}

\subsection{Real-World Attack Feasibility (RQ3)}
\noindent\textbf{Targets.} We study two target surfaces.
\textit{Off-chain targets:} log consumers, including blockchain explorers, wallets, NFT
marketplaces, indexers, and bridge relayers. These are selected from (1) previously reported
sleepminting incidents~\cite{sleepminting}, (2) wallet programs listed on bug-bounty platforms
such as BugRap~\cite{bugrap}, and (3) widely used blockchain explorers (e.g., BscScan, OKLink,
Bitquery, Tokenview, and Bsctrace)~\cite{bscscan,oklink,bitquery,tokenview,bsctrace}. For EOCA on
bridges, we additionally search open-source bridge projects and historically affected bridge
relayers with public audit reports (e.g., Sygma relayer audits)~\cite{sygma_relayer}.
\textit{On-chain targets:} deployed contracts whose events are used as off-chain triggers, drawn
from a set of 100 active contracts listed on the Dune dashboard ``How Users React''~\cite{userreact}.

\noindent\textbf{Methodology.} We separate targets into two groups: (i) off-chain consumers (e.g.,
explorers, wallets, relayers) for testing the two attack vectors and (ii) on-chain contracts for
UEM auditing. For off-chain targets, we collect test addresses/contracts and derive candidate
scenarios for the two attack vectors. Using our off-chain harness, we craft and send test
transactions (on a forked local chain or testnet) and then verify whether emitted logs and state
transitions satisfy emitter constraints, event signatures, indexed topics, and state deltas. We
also observe downstream behaviors to determine whether the events are accepted or misinterpreted.
When public evidence is incomplete, we reproduce the workflow in a controlled environment or
replay historical attack transactions, avoiding any impact on live systems. For on-chain targets,
we run \tool over the 100 active contracts from the Dune dashboard to identify unauthorized event
emission and manually validate candidates. We responsibly report confirmed cases.

\noindent\textbf{Result.} \textbf{(1) EOCA.} We classify EOCA as cases where off-chain consumers accept
events from unintended emitters due to insufficient emitter validation, and we confirm emitting
contracts differ from legitimate project addresses. We identify some bridge-relayer pipelines
vulnerable to emitter confusion; we reproduce one historically affected bridge relayer and confirm
a newly affected open-source bridge relayer implementation that has been forked and deployed, with
reports filed and acknowledged~\cite{sygma_relayer,bridgecore_dl_tokene,bridgecore_qtum}.
\textbf{(2) ESDA.} We classify the following cases as ESDA because the off-chain consumer infers
transfers purely from event logs without validating corresponding state changes, and we confirm
missing balance/ownership deltas on-chain: (a) We observe transfer-spoofing transactions that emit
transfer logs without corresponding token state changes, causing apparent movements from
\texttt{0x8888...8888} to a victim wallet; despite no actual token movement, at least five major
explorers (BscScan, OKLink, Bitquery, Tokenview, and Bsctrace) display the spoofed event as
legitimate~\cite{bscscan,oklink,bitquery,tokenview,bsctrace}; (b) During audits, we found ESDA
issues in six cryptocurrency wallets, reported them, and received four confirmations, including a
\$600 bounty; two reports remain pending; (c) We identified a log-display issue in Blockscout that
can mislead users about event data~\cite{errordisplay}; (d) We design a bypass to reproduce the
sleepminting pattern on testnets for two major NFT marketplaces (OpenSea~\cite{OpenSea} and
Rarible~\cite{Rarible}).
\textbf{(3) UEM.} In our on-chain auditing, we find three DeFi projects and one GameFi project
vulnerable to unauthorized event emission (privileged events emitted without proper access
control); the largest affected project had a market capitalization of \$169{,}688 as of
Oct.~15,~2024, based on Etherscan~\cite{etherscan}.

\textbf{Finding: Insufficient event semantic validation creates on-chain/off-chain inconsistencies.}
Our results reveal three root causes: (1) off-chain services that neglect emitter validation can be tricked into processing events from attacker-controlled contracts, enabling unauthorized operations; (2) off-chain services that omit state-delta checks display spoofed transfers as legitimate, misleading users about actual token holdings; and (3) contracts lacking access control for privileged events allow attackers to emit arbitrary logs, causing off-chain services to record fabricated data that diverges from true on-chain state.

\subsection{Threats to Validity}
\label{sec:tov}
\revised{
\noindent\textbf{Internal validity.} (1)~\emph{Data extraction and labelling:} Manual filtering, taxonomy construction, and ground-truth labelling may involve subjective judgement. We mitigate this through independent double labelling, inter-rater agreement reporting, disagreement resolution, and validation by three industry smart-contract security experts. (2)~\emph{Severity treatment:} We do not stratify findings by severity because impact depends on off-chain log interpretation; a low-severity audit finding can have critical impact when consumed by a vulnerable bridge or wallet. This affects severity analysis only. (3)~\emph{Detection accuracy:} Detection accuracy is bounded by decompilation failures and solver/time limits. We manually cross-validated all 95 labelled samples to catch decompilation-induced false negatives at labelling time. (4)~\emph{Baseline comparability:} Because general-purpose analyzers are not designed for event semantics, we conservatively map detectors to categories and exclude non-event detectors.
}

\revised{
\noindent\textbf{External validity.} Generalisability depends on how well the contract corpus and taxonomy-source dataset represent real-world practice. (1)~\emph{Data representativeness:} The contract corpus is drawn from Smart Contract Sanctuary~\cite{smartcontractsanctuary}, whereas the taxonomy-source dataset covers FORGE-derived~\cite{forgeicse2026} audit findings and confirmed security incidents reported through the end of 2025. Public incident reports and audit findings may still underrepresent rare or undocumented event-semantic issues. (2)~\emph{Evaluation scope:} Our evaluation targets active contracts on a pool of \TargetPoolSize contracts, and off-chain case studies cover bridges, explorers, wallets, and marketplaces; results may not generalise to inactive or highly specialised contracts. (3)~\emph{Temporal stability:} Finalised ERC event signatures are immutable after standardisation, so inferred specifications remain applicable to newer contracts that implement those standards. (4)~\emph{Corpus independence:} The \tool pipeline does not hard-code corpus-specific assumptions, so it can be re-run on any sufficiently large bytecode corpus to produce updated specifications without changing the methodology, taxonomy, or detection rules.
}

%% file: Sections/6_Related.tex
\section{Related Work}
\label{sec:related}
\subsection{Smart Contract Security Analysis}
Smart-contract analyzers span several paradigms. \textbf{Static analysis} includes XML/XPath
rules in SmartCheck~\cite{tikhomirov2018smartcheck}, bytecode decompilation in
Gigahorse~\cite{grech2022elipmoc}, source-level rules in Slither~\cite{Slither}, AST analysis in
NeuCheck~\cite{lu2021neucheck}, and bytecode taint tracking in eTainter~\cite{ghaleb2022etainter}.
\textbf{Symbolic execution} ranges from Oyente and source-level SolSEE~\cite{luu2016making,lin2022solsee}
to inter-contract or slicing-based systems such as Pluto, CRUSH, and
JACKAL~\cite{ma2021pluto,ruaro2024not,gritti2023confusum}; Mythril and Manticore are widely used
implementations~\cite{mythril,mossberg2019manticore}. \textbf{Dynamic analysis} includes
ContractFuzzer, ConFuzzius, and Smartian~\cite{contractfuzzer2018,confuzzius2021,smartian2021},
while \textbf{formal verification} systems such as Zeus, Securify, and Maian check specified
properties~\cite{zeus2018,securify2018,maian2018,specmin}.

These systems target general vulnerability templates or user-supplied properties, including
reentrancy, access control, arithmetic, storage, and call-flow properties, rather than event
semantics. Event-semantic checking requires a reference for emitter, timing, and parameter--state
consistency. Formal approaches require that reference in advance, whereas \tool infers event and
function profiles from deployed bytecode and differentially checks targets, including non-ERC
events and conventions without written specifications.

\subsection{Smart Contract Events}
Prior work studies event usage and gas-inefficient emissions~\cite{li2023understanding}, cross-chain
security events and vulnerabilities~\cite{wu2026traphunter,Xscope,liao2024smartaxe,wang2024xguard}, log-based token
tracking~\cite{cernera2023token,Liu2025DRT}, sleepminting~\cite{sleepminting,xiao2025wakemint},
code--documentation--event inconsistencies~\cite{doccon}, and ERC-20 behavioral mismatches~\cite{TokenScope}.
These studies focus on particular domains, while off-chain infrastructure still reconstructs state
from logs and ABI expectations without necessarily validating emitter or on-chain state. Our work
provides a cross-domain taxonomy of event-semantic defects and recurring off-chain attack vectors
that explicitly covers emitter and state--event consistency.

%% file: Sections/7_Discussion.tex
\section{Discussion and Mitigation}
\label{sec:mitigation}

Event-semantic issues arise from both contract logic and off-chain interpretation. At the
\textbf{contract level}, developers should emit events after state updates, guard privileged
emissions, avoid ambiguous signature reuse, validate ambiguous inputs (e.g., zero-address versus
native-token semantics), and keep parameter order and indexing consistent with the ABI. At the
\textbf{off-chain level}, consumers should allowlist emitters; validate signatures, topic counts,
parameter types, and reverted status; and confirm state or call context for high-risk transfers,
minting, and withdrawals. Consumers should also bind each event to its intended call and validate
the complete sequence (count, order, emitter, and call path), while monitoring log signals against
state invariants and versioned schemas.

\revised{
\noindent\textbf{Quantitative overhead of on-chain state verification.} We traced 26 Ethereum
mainnet transactions from four Xscope cross-chain incidents (THORChain \#1/\#2/\#3 and Qubit
Bridge) involving Inconsistent Event Parsing or Unrestricted Deposit Emitting~\cite{xscope-results}.
For each, we called Tenderly's \texttt{debug\_traceTransaction} with the \texttt{prestateTracer}
(diffMode) three times and averaged the round-trip latency~\cite{tenderly}. In deployment, address
allowlists and event signatures can filter candidates before tracing.

Across all 26 transactions, the median trace latency is \SI{900}{\milli\second} (average
\SI{1381}{\milli\second}, P95 \SI{3030}{\milli\second}, max \SI{7238}{\milli\second}). All traces complete well
within Ethereum's \SI{12}{\second} block budget; the slowest uses 60\% of it. The tracer returns
pre/post storage differences for checking event--state correspondence, and a local archive node
can remove network round trips. The results support transaction tracing for real-time verification.
}

%% file: Sections/8_Conclusion.tex
\section{Conclusion}

\tool detects event-semantic defects through corpus-inferred event and function profiles and
explainable differential checking. We define five defect types and two off-chain attack vectors.
Across \TargetPoolSize contracts and labeled data, \tool achieves 90.17\% comprehensive precision.
Its harness reproduces both vectors and confirms issues in relayers, explorers, wallets, and NFT
marketplaces. We also provide mitigations for contracts and off-chain consumers.

%% file: Sections/9_DataAvailability.tex
\section*{Data Availability}
\label{Sec:DataAvailability}
Our replication package is available online at
\url{https://github.com/yxsec/eventspec}.